\documentclass[english]{article}
\usepackage[T1]{fontenc}
\usepackage[latin9]{inputenc}
\usepackage{amsmath}
\usepackage{amssymb}
\PassOptionsToPackage{normalem}{ulem}
\usepackage{ulem}
\usepackage{graphicx}
\usepackage{url}

\makeatletter

\newcommand{\noun}[1]{\textsc{#1}}
\providecommand{\tabularnewline}{\\}
\usepackage{babel}
\makeatother
\begin{document}

\title{A GPU-Based Genetic Algorithm for the P-Median Problem}

\author{\noun{Bader F. AlBdaiwi}
and \noun{Hosam M.F. AboElFotoh}
\\ \\
{Computer Science Department, Kuwait University, Kuwait}
\\ \\
{\normalsize\{bdaiwi,\ hosam\}@cs.ku.edu.kw}
}
\maketitle
\begin{abstract}
The p-median problem is a well-known NP-hard problem. Many heuristics
have been proposed in the literature for this problem. In this paper,
we exploit a GPGPU parallel computing platform to present a new genetic
algorithm implemented in Cuda and based on a Pseudo Boolean formulation
of the p-median problem.
We have tested the effectiveness of our algorithm using a Tesla K40 (2880
Cuda cores) on 290 different benchmark instances obtained from OR-Library,
discrete location problems benchmark library, and benchmarks introduced
in recent publications.
The algorithm succeeded in finding optimal solutions for all instances
except for two OR-library instances, namely pmed30 and pmed40, where
better than 99.9\% approximations were obtained.
\end{abstract}
\begin{description}
\item [{Keywords}] P-Median Problem; NP-Hard; GPGPU; Cuda; Pseudo Boolean
Formulation; Genetic Algorithms; Heuristics. 
\end{description}

\section{Introduction}
The P-Median Problem (PMP) is formally defined as follows. Given a
set $C=\{1,\ldots,n\}$ of $n$ clients, a set $F=\{1,\ldots,m\}$
of $m$ facilities, an integer $p<m$, and the distance $d_{ij}$
between client $i$, $1\leq i\leq n$, and facility $j$, $1\leq j\leq m$.
Let $y_{ij}\in\{0,1\}$ be a decision variable such that $y_{ij}=1$
if and only if client $i$ ~is serviced by facility $j$, and let
$x_{j}\in\{0,1\}$ be a decision variable such that 
$x_{j}=1$ if and only if facility $j$ is open for service.
The PMP objective is to minimize the total
distance\footnote{We shall use distance and cost interchangeably}
\begin{equation}
f_{C}(x,y)=\sum_{i\epsilon C}\sum_{j\epsilon F}x_{j}y_{ij}d_{ij},
\end{equation}

subject to 
\begin{equation}
\sum_{j\in F}y_{ij}=1\text{\ensuremath{\,\,\,\,\forall i\in C}},
\end{equation}

\begin{equation}
\sum_{j\epsilon F}x_{j}=p.
\end{equation}

The objective function (1) minimizes the total distance between clients
and the corresponding service facilities. Constraint (2) states that
each client is serviced by exactly one facility. Constraint (3) states
that the number of open facilities is exactly $p$. Let the set of
open facilities be $O=\{o_{1},\ldots,o_{p}\}$. Naturally, if client
$i$ is serviced by facility $j$, ($y_{ij}=1$ ), then: 1) $j \in O$
(is open), and 2) $d_{ij}$ is a minimum over \{ $d_{io_{1}}...d_{io_{p}}$\}.
An instance of the PMP is described by an $n\times m$ distance matrix
$C=[d_{ij}]$ and a positive integer $p<m$.
Note that we assume the elements of $C$ are non-negative.

The PMP has a wide range of applications. It has been extensively
researched in the literature. It has many applications in logistics
~\cite{bargos2016location}\cite{jaillet1996airline} and location
science~\cite{laporte2015location}\cite{ren2015investigating}.
It also has applications in finance and market analysis~\cite{goldengorin2014pseudo}.
Unfortunately, it is NP-hard and hence difficult to solve it for optimality~\cite{kariv1979}.
Comprehensive surveys on solving methods for the PMP and its variations
can be referred to in~\cite{Daskin2015}\cite{farahani2013hub}\cite{mladenovic2007p}\cite{reese2006solution}.

In this paper, we exploit a GPGPU parallel computing platform to present
a new genetic algorithm implemented in Cuda C version~7.5 (Compute Unified Device Architecture)
and based on a Pseudo Boolean formulation of the PMP.

The rest of the paper is organized as follows.
Section~2 introduces preliminaries and related literature review on the pseudo
boolean formulation of the PMP, GPGPU and Cuda, and genetic algorithms.
Section~3 presents the new algorithm. Section~4 highlights some
implementation details. The algorithm time complexity is analyzed
in Section~5. Section~6 presents the experimentation results, and Section~7
concludes the paper.

\section{Preliminaries}

\subsection{A Pseudo Boolean Formulation of the PMP}

The pseudo Boolean formulation of the PMP appeared in~\cite{Hammer}.
It is obtained as follows. For each client $i,$ let $\prod^{i}=(\pi_{i1},\ldots,\pi_{im})$
be an ordering of $1,\ldots,m$ such that $d_{i\pi_{ik}}\leq d_{i\pi_{il}}$
if $k<l$ for all $k,l\in\{1,\ldots,m\}$, and let $\triangle^{i}=(\delta_{i1},\ldots,\delta_{im})$,
where $\delta_{i1}=d_{i\pi_{i1}}$, and $\delta_{ir}=d_{i\pi_{ir}}-d_{i\pi_{i(r-1)}},$
for $r=2,...,m$. Therefore, the distance between client $i,$ $i\in\{1,\ldots n\}$,
and the facility serving it can be expressed using the following pseudo Boolean
polynomial 
\begin{equation}
d_{i}=\delta_{i1}+\sum_{k=2}^{m}\delta_{ik}\prod_{r=1}^{k-1}\bar{x}_{\pi_{ir}}\ .\label{eq:4}
\end{equation}
Thus, the PMP can be reformulated as: Given $\triangle=[\delta_{ij}]$,
$\prod=[\pi_{ij}]$, and $p$, find an assignment in \{0, 1\} to $x_{i},$
$i\in\{1,\ldots,n\}$, such that 
\begin{equation}
\sum_{j\in F}x_{j}=p,
\end{equation}

and

\begin{equation}
B_{C}(z)=\sum_{i=1}^{n}\left(\delta_{i1}+\sum_{k=2}^{m}\delta_{ik}\prod_{r=1}^{k-1}z_{\pi_{ir}}\right)
\end{equation}
is minimized. The Boolean variable $z_{j}=\bar{x_{j}}$ is 1 iff $x_{j}=0$,
denoting a closed facility. Note that the $k^{th}$ term in Equation~(\ref{eq:4})
contains $\prod_{r=1}^{k-1}\bar{x}_{\pi_{ir}}$. Therefore, this term
must be zero $\forall\ k>m-p+1$ since at least one facility (say
$f$) is open in any $m-p+1$ locations resulting in $\bar{x}_{f}=0$.
Thus, $\prod$ and $\triangle$ can be reduced to $\prod^{'}$ and
$\triangle^{'}$ by omitting the last $p-1$ columns.

The objective function based on $\prod^{'}$ and $\triangle^{'}$
is known as \emph{Hammer-Bersnev polynomial} (\emph{HBP})\cite{albdaiwi2009equivalent}.
It can be further reduced through monomial reduction. Interested readers
could refer to~\cite{BaderAlBdaiwi2011} for details. Example~1 illustrates the PMP
pseudo boolean formulation.
\\

\noindent
\emph{Example 1 }: Consider a PMP instance with $n=5$, $m=4$, $p=2$
and 
\[
C=\left[\begin{array}{cccc}
7 & 10 & 16 & 11\\
15 & 17 & 7 & 7\\
10 & 4 & 6 & 6\\
7 & 11 & 18 & 12\\
10 & 22 & 14 & 8
\end{array}\right].
\]
\\
An ordering matrix $\prod$ and the corresponding matrix $\triangle$
are given by 

\begin{center}
$\prod=\left[\begin{array}{cccc}
1 & 2 & 4 & 3\\
3 & 4 & 1 & 2\\
2 & 3 & 4 & 1\\
1 & 2 & 4 & 3\\
4 & 1 & 3 & 2
\end{array}\right],
\
\triangle=\left[\begin{array}{cccc}
7 & 3 & 1 & 5\\
7 & 0 & 8 & 2\\
4 & 2 & 0 & 4\\
7 & 4 & 1 & 6\\
8 & 2 & 4 & 8
\end{array}\right].$ 
\par\end{center}
Omitting the last $(p-1=1)$ column corresponding to zero terms in
Equation~(\ref{eq:4}) results in:\\

\begin{center}
$\prod^{'}=\left[\begin{array}{cccc}
1 & 2 & 4\\
3 & 4 & 1\\
2 & 3 & 4\\
1 & 2 & 4\\
4 & 1 & 3
\end{array}\right],$ $\triangle^{'}=\left[\begin{array}{cccc}
7 & 3 & 1\\
7 & 0 & 8\\
4 & 2 & 0\\
7 & 4 & 1\\
8 & 2 & 4
\end{array}\right].$ 
\par\end{center}
The corresponding HBP representing total distance (cost) is 
\[
B_{C}(z)=\begin{array}{l}
\left[7+3z_{1}+1z_{1}z_{2}\right]+\left[7+0z_{3}+8z_{3}z_{4}\right]+\left[4+2z_{2}+0z_{2}z_{3}\right]+\\
\left[7+4z_{1}+1z_{1}z_{2}\right]+\left[8+2z_{4}+4z_{1}z_{4}\right].
\end{array}
\]

$B_{C}(z)$ has $n\times(m-p+1)=15$ entries, and the reduced polynomial
is 
\begin{eqnarray*}
B_{C}(z) & = & 33+7z_{1}+2z_{2}+2z_{4}+2z_{1}z_{2}+8z_{3}z_{4}+4z_{1}z_{4}.
\end{eqnarray*}

\subsection{GPGPU and Cuda}

Similar to many NP-hard problems, many heuristics have been developed
for the PMP. Some of these heuristics tried to exploit parallel computing
platforms to reach a near-optimal solution in a reasonably short time~\cite{Daskin2015}.
A few years ago, Nvidia has introduced Cuda (Compute Unified Device
Architecture) that provides an application interface (API) for general
purpose computing~\cite{Cuda C}. Hence, GPGPU refers to General
Purpose computing on Graphics Processing Units. Nvidia graphics cards
(as well as all graphics cards) are designed to do similar computations
on large numbers of pixels. Therefore, they contain hundreds of processing
elements (cores), although not as powerful as CPU cores, that execute
thousands of similar threads (grouped in blocks) in parallel. Unlike
CPU's threads, the context switching between blocks of threads requires
a minimal overhead. There are not many GPU-based solutions for the
PMP so far. Lim and Ma introduced GPU implementations for solving
the PMP using the vertex substitution and Llod algorithms in~\cite{Lim2013}
and~\cite{ma2011gpu}. Cuda C can be used for developing applications
on GPGPU. In our implementation, we used Cuda C version~7.5.

A Cuda C program consists of two types of code: \emph{Host}
code and \emph{Device} code. The Host code refers to that executed
by the CPU and the device code refers to that executed by the GPU
card. The Host code launches a \emph{kernel} that is executed by the
device. A kernel launch specifies the number of threads to be executed in
parallel. These threads are grouped in blocks. Blocks in their turns
are organized in a grid. All threads execute the same code but on
multiple data, which Nvidia calls a \emph{Single Instruction Multiple
Thread} (SIMT) architecture. To support multidimensional data modeling
and processing, Cuda enables defining grids and blocks to be single,
double, or triple dimensions. For example, the following Cuda code
declares two triple dimensional arrays using the \emph{dim3} data
type. Then, it invokes \emph{KernelX} with a $3\times4\times6$ grid
each element of which is a $2\times4\times4$ block.

\[
\begin{array}{l}
dim3\ Grid(3,4,6);\\
dim3\ Block(2,4,4);\\
KernelX<<<Grid,Block>>>(\mathit{/*}\ Parameter\ List\ \mathit{*/});
\end{array}
\]
The total number of blocks in \emph{KernelX} is $(3\times4\times6=72)$,
and each block has $(2\times4\times4=32)$ threads. Thus, the total
number of threads in \emph{KernelX} is $(72\times32=2,304)$. To differentiate
among threads, Cuda defines two built-in variables for block and thread
indexing, namely \emph{blockIdx} and \emph{threadIdx}. Each of these
variables is \emph{3-dimensional}. For instance, \emph{threadIdx}
has three components \emph{threadIdx.x,} \emph{threadIdx.y}, and \emph{threadIdx.z}.
Cuda also defines two other built-in \emph{3-dimensional} variables
for the grid and block dimensions, \emph{gridDim} and \emph{blockDim}.
They are automatically initialized at a kernel launch. In the above
example, \emph{KernelX} sets \emph{gridDim.x = 3}, \emph{gridDim.y
= 4}, and \emph{gridDim.z = 6}. It also sets \emph{blockDim.x = 2},
\emph{blockDim.y = 4}, and \emph{blockDim.z = 4}.

Linearized unique block identifier, \emph{BID}, and linearized unique
global thread identifier, \emph{TID}, can be derived from \emph{gridDim},
\emph{blockDim}, \emph{blockIdx}, and \emph{threadIdx} as follows:
\[
\begin{array}{ccl}
BID & = & blockIdx.x+blockIdx.y*gridDim.x+\\
 &  & gridDim.x*gridDim.y*blockIdx.z;\\
\\
TID & = & BID*(blockDim.x*blockDim.y*blockDim.z)+\\
 &  & (threadIdx.z*(blockDim.x*blockDim.y))+\\
 &  & (threadIdx.y*blockDim.x)+threadIdx.x;
\end{array}
\]

Threads may access data in parallel from different memory locations.
Usually, the data accessed by each thread is determined by its index (within
the block) or its global identifier. In Cuda, the device code can
access only the device memory. Therefore, the host code has to initialize
the device memory through Cuda calls that allocate device memory (\emph{cudaMalloc})
and copy (\emph{cudaMemcpy}) data between host RAM and device RAM.
The device memory has three basic types: 
\begin{description}
\item [{Global}] can be accessed by all threads from all blocks. 
\item [{Shared}] can be accessed by all threads in the block. Each block
has a limited amount of shared (on-chip) memory. 
\item [{Private}] can be accessed only by the thread itself. Each thread is allocated
a limited number of registers. 
\end{description}
The global memory is too slow compared to the shared and private memories.
Therefore, the shared and private memory have to be utilized to the maximum
extent.

\subsection{Genetic Algorithms}

One of the most suitable heuristics framework (\emph{Meta Heuristics})
that exploits the availability of many cores performing the same instruction
thread on multiple data is \textit{Genetic Algorithms} (GA)~\cite{Mitchell1998}.
Recently, efficient GPU-based GA are being proposed for solving hard
problems. For example, Kang et. al. have introduced such a solution
for the Traveller Salesman Problem (TSP) in~\cite{Kang2016}. We
have not encountered any GPU-based GA for the PMP; even though, GA
are recognized as one of the most effective evolutionary technique
for solving optimization problems.

In GA, a large number (\emph{Population}) of \textit{Chromosomes} are generated and
operated upon using similar operations like \textit{mutations}, \textit{crossover,
migration} and \textit{fitness test}. A chromosome is a finite sequence
of genes commonly represented by a binary string or a set of integers.
A crossover operation involves two parent chromosomes exchanging genes
to produce offsprings, while a mutation involves only a single chromosome
that mutates into a new one. Usually, each chromosome represents a
candidate solution to the optimization problem, and the fitness of
the chromosome represents the solution quality or the objective function
value.

There exists a number of GA for solving the PMP in the literature,
examples are~\cite{Alp2003}\cite{bozkaya2002efficient}\cite{jaramillo2002use}.
Combining GA with local search method results in hybrid GA that could
speed up the reach for global optima~\cite{el2006hybrid}. Hybrid
GA for solving the PMP appear in~\cite{rebreyend2015computational}\cite{resende2004hybrid}.
Hybrid GA based on the variable neighborhood search have been recently
published in~\cite{drezner2015new}\cite{todosijevic2015general}.
Different hybrid GA for the PMP using solution archiving, greedy strategy,
and fine-grained tournament selection are presented in~\cite{biesinger2015hybrid}\cite{kazakovtsev2015modied}
and~\cite{stanimirovic2012genetic}, respectively.

\section{A New Genetic Algorithm for the PMP Based on GPU and Pseudo Boolean
Formulation}

The algorithm is quite simple.
The host randomly generates an initial
population of chromosomes (candidate solutions) and passes them to
a device kernel for fitness evaluations and enhancements.
This basically iterates
with migrating the best fit chromosomes from the current population
to the next. The algorithm terminates at reaching an iteration limit
or over saturating the solution enhancement.

Parallelism manifests in our algorithm in two folds.
First, the host generates the next population in parallel
with the device processing of the current
population fitness evaluations and enhancements.
Second, the kernel threads run in parallel, and
each of which is assigned a chromosome for
fitness evaluation and enhancement.
The fitness evaluation is based on $\prod^{'}$ and $\triangle^{'}$
whose matrix structure harnessed the PMP data parallelism potentials
as explained in~\ref{fitness}.
A chromosome enhancement is based on crossover and mutation operations.
The details of these operations are explained
in Subsections~\ref{crossover} and~\ref{Mutation}.

The following two subsections outline the algorithm
host and device codes.

\subsection{Host Code}

 $n$: Number of Clients,\\
 $m$: Number of Facility Locations,\\
 $p$: Number of Open Facilities,\\
 $C$: $n\times m$ Distance/Cost Matrix\\
 $\mathit{NB}$: Number of GPU Blocks\\
 $NT$: Number of Threads per Block\\
 $EvolveLimit$: Limit on Number of Calls to \emph{Evolve} Kernel\\
 $S$: Saturation Limit\\

\noindent \textbf{Steps:} 
\begin{enumerate}
\item Read input \emph{n, m, p, C, NB, NT, KernelsLimit}. 
\item Call kernel (\emph{Init} <\textcompwordmark{}<\textcompwordmark{}<
\emph{NB, NT} >\textcompwordmark{}>\textcompwordmark{}>) to differently
seed \emph{curand} in each thread.
\item Compute $\prod$ and $\triangle$ matrices each of size $\left(n\times m\right)$,
and reduce them to $\prod^{'}$ and $\triangle^{'}$ each of size
$\left(n\times m-p+1\right)$. 
\item Allocate device memory for $\prod^{'}$ and $\triangle^{'}$ using
\emph{cudaMalloc}. 
\item Copy $\prod^{'}$ and $\triangle^{'}$ to device memory using \emph{cudaMemcpy}.
\item Initialize a counter for the Number of \emph{Evolve} kernal calls:
NKernels = 0.
\item Generate candidate solutions as a random \emph{population} of $\mathit{NB}\times NT$
\emph{chromosomes}. 
\item Wait for \emph{Init} kernel to finish (\emph{cudaDeviceSynchronize}). 
\item Copy current population to device memory (\emph{cudaMemcpy}). 
\item Call kernel (\emph{Evolve} <\textcompwordmark{}<\textcompwordmark{}<
\emph{NB, NT} >\textcompwordmark{}>\textcompwordmark{}>). 
\item Generate a new random population of $\mathit{NB}\times NT$ \emph{chromosomes}
for next \emph{Evolve} kernel call. 
\item Wait for \emph{Evolve} kernel to finish (\emph{cudaDeviceSynchronize}).
\item Copy \emph{most fit chromosome }(best solution) of each block to the
Host memory and find \emph{their most fit}. 
\item Increment NKernels. 
\item If (the \emph{most fit chromosome} has not been changed by the last
$S$ \emph{Evolve} kernal call) or (NKernels >= \emph{EvolveLimit}),
report the \emph{most fit chromosome} as the best solution and stop; 
\item Else \emph{Migrate} \emph{NB} best chromosomes to next population
and \emph{Go to} Step 9. 
\item END (Host Code). 
\end{enumerate}

\subsection{Device Code (Executed in Parallel by Each Thread)}

\subsubsection{Init Kernel}

\noindent \textbf{Steps:} 
\begin{enumerate}
\item Call \emph{curand\_init(0, TID, 0, \&state)},\\
(Thread Global Identifier) is passed as a \emph{sequence number}~\cite{NVIDIA2015}.
This insures each thread to have a different random sequence when
calling \emph{curand}. 
\item END (Init Kernel). 
\end{enumerate}

\subsubsection{Evolve Kernel}

\paragraph{Global Memory:}
\begin{itemize}
\item Array B\_MinCost{[}NB{]}: Minimum cost (Highest fitness) found by
each block. 
\item Array Best{[}NB{]}: Chromosome with best fitness value for each block. 
\end{itemize}
\textbf{Shared Memory: } 
\begin{itemize}
\item Array MinCost{[}NT{]} : Minimum cost found by each thread in the block,
initially set to MinCost{[}0{]}. 
\item txMin{[}NT{]}: used to find the Index of the thread with best fitness
value in the block, initially set to threadIdx.x. 
\end{itemize}

\paragraph{Steps:}
\begin{enumerate}
\item Evaluate the fitness of the thread chromosome $\mathbb{C}$ using
$\prod^{'}$ and $\triangle^{'}$ matrices. 
\item Initialize relative thread index and relative block size: rtx = threadIdx.x;
rb\_size = NT. 
\item Crossover Cycle:\\
 For ( CStride = NT /2 ; CStride >0; CStride = CStride/2) 

\begin{enumerate}
\item Generate a random crossover point r1 (using \emph{curand}). %
\item If (rtx >= rb\_size/2), Stride = -CStride else Stride = CStride. %
\item Form the parent couples by finding a unique couple for each thread:
Couple = TID + Stride; 
\item Make \emph{cross-over} between $\mathbb{C}$ and Couple at r1 and
form offspring $\mathcal{F}$. 
\item If (Fitness($\mathcal{F}$) < Fitness($\mathbb{C}$)), replace $\mathbb{C}$
by $\mathcal{F}$. 
\item rb\_size = rb\_size /2; rtx = rtx \% rb\_size; 
\end{enumerate}
\item Synchronize each block threads using \emph{syncthreads()}. 
\item Mutation Cycle:\\
 For (i= $lg$(NT), Enhanced = False; (i > 0) and (Not Enhanced);
i = i-1) 

\begin{enumerate}
\item Randomly decide the mutation parameters as per the details explained
in~\ref{Mutation}. 
\item Mutate $\mathbb{C}$ to offspring $\mathcal{F}$. 
\item If Fitness($\mathcal{F}$) < Fitness($\mathbb{C}$), 

\begin{enumerate}
\item Replace $\mathbb{C}$ by $\mathcal{F}$. 
\item Enhanced = True. 
\end{enumerate}
\end{enumerate}
\item Synchronize each block threads using \emph{syncthreads()}. 
\item Find best fitness in the block:\\
 For ( Stride = NT /2 ; Stride >0; Stride = Stride/2) 

\begin{enumerate}
\item tx = threadIdx.x. 
\item If (tx < Stride and (MinCost{[}tx+Stride{]} < MinCost{[}tx{]})) 

\begin{enumerate}
\item MinCost{[}tx{]} =MinCost{[}tx+Stride{]}. 
\item txMin{[}tx{]}=txMin{[}tx+Stride{]}. 
\end{enumerate}
\end{enumerate}
\item If (threadIdx = 0), Store each block best fitness cost and its corresponding
chromosome in B\_MinCost{[}blockIdx.x{]} and Best{[}blockIdx.x{]},
respectively. 
\item End (Evolve Kernel). 
\end{enumerate}

\section{Implementation Details}

\subsection{Chromosome Representation and Generation}
\label{generation}

A chromosome is represented as a vector $\mathbb{C}$ of $m$ bits
$\mathbb{C}_{0}$:$\mathbb{C}_{m-1}$, where \emph{true} denotes an
open facility and \emph{false} denotes a closed one. Our algorithm
generates chromosomes by random selection from a lexicographical order
of a combinatorial sequence~\cite{hall1965combinatorial}. For each
chromosome, it first generates a non-negative integer $i<\binom{m}{p}$
using a 64-bit random function. Next, it generates the $i^{th}$ Lexicographic
combination of $\binom{m}{p}$ using the efficient method presented
in~\cite{McCaffreyJuly2004}. As a result, the generated chromosome
will have exactly $p$ true bits each of which corresponds to a selected
element in the $i^{th}$ Lexicographic combination of $\binom{m}{p}$.
This method reduces the random function calls to one call per chromosome
generation. Hence, it could maintain better quality random number
generation as it limits the probability of exhaustively consuming
the pseudo random sequence generated by the utilized random function.
It has positively impacted the quality of the solutions generated
by our algorithm. Interestingly, we are not aware of its existence
in the literature.

\subsection{Fitness Function}

\label{fitness}

The Fitness function is a performance bottleneck. Each thread calls
it several times. It is called to evaluate a thread assigned chromosome.
It is also called in each crossover iteration to evaluate the offsprings.
Furthermore, it is called to evaluate the mutation offsprings.
Therefore,
we designed this function to be as efficient as possible. In general,
we harnessed the data parallelism in the PMP by using the pseudo boolean
formulation and tailored it to be GPU suitable. We designed the fitness
function to use $\prod^{'}$ and $\triangle^{'}$ rather than HBP.
This enabled higher degree of data parallelism and restricted the
required operations to be simple integer additions. We also took advantage
of memory caching when accessing the $\prod^{'}$ and $\triangle^{'}$
simply because all threads in all blocks use exactly the same $\prod^{'}$
and $\triangle^{'}$ in read only mode.

The function algorithm is straightforward. For an input chromosome $\mathbb{C}$, 
it scans the entry of each client in $\triangle^{'}$ as per the order of $\prod^{'}$
and accumulates the corresponding increments till an open facility is found
(\emph{$\mathbb{C_{S}=}$ true}).
The total accumulations of all clients represents
the fitness of $\mathbb{C}$.

\paragraph{function Fitness ( $\mathbb{C}:$ Chromosome ) }
\begin{enumerate}
\item fitness = 0; 
\item \emph{for} \emph{each} client $i$ , $1\leq i$ $\leq$ n: 
\begin{itemize}
\item \emph{S =0;}
\item \emph{repeat}: \emph{S++}; ~fitness = fitness + $\delta_{i\pi_{is}}
       $~~\emph{until} $\mathbb{C_{S}}$
\end{itemize}
\item End (Fitness). 
\end{enumerate}

The Fitness function time order is $O(n(m-p))$ since each
of $\prod^{'}$ and $\triangle^{'}$ is of size $n\times(m-p+1)$.
In average, there will be one open facility in each~$\frac{m-p-1}{p} \approx \frac{m}{p}$
locations assuming the~$p$ open facilities are normally distributed.
Therefore, the Fitness function expected average time is
$O(\frac{nm}{p})$.
Based on this, we decided to compute each chromosome fitness by a single
thread accumulating the increments of all clients.
This results in a better device core utilization and higher scalability
as explained in~Section~\ref{Experimentation}.
Consequently, we designed the Evolve kernel grid and blocks
to be of single dimensions.
\subsection{Crossover Operation}
\label{crossover}

Each thread determines its unique couple as in Step 3 of the 
Evolve kernel. One of the couple threads generates two random
integers and shares them with its couple.
The first integer $r1$, $0 \le r1 < m$, determines the
crossover starting index.
The second integer $r2 = 2i$, $ 0 < i \le \lfloor\frac{p}{2}\rfloor$,
determines the number of genes to be exchanged.
The exchanges count and occur only between unequal corresponding genes.
In order to keep exactly~$p$ true genes in the offspring,
exactly $i$~genes are exchanged from~$0$ to~$1$, and
the other remaining $i$~genes are exchanged from~$1$ to~$0$.
If the end of the chromosome
is reached before having the right number of exchanges, the search
continues from the beginning.
If the the number of exchanges
cannot reach~$r2$, the operation fails and no offspring is produced.
Figure~1
shows a crossover operation example.
\begin{figure}
\label{Fig1}
\centering
\includegraphics[width=0.75\textwidth]{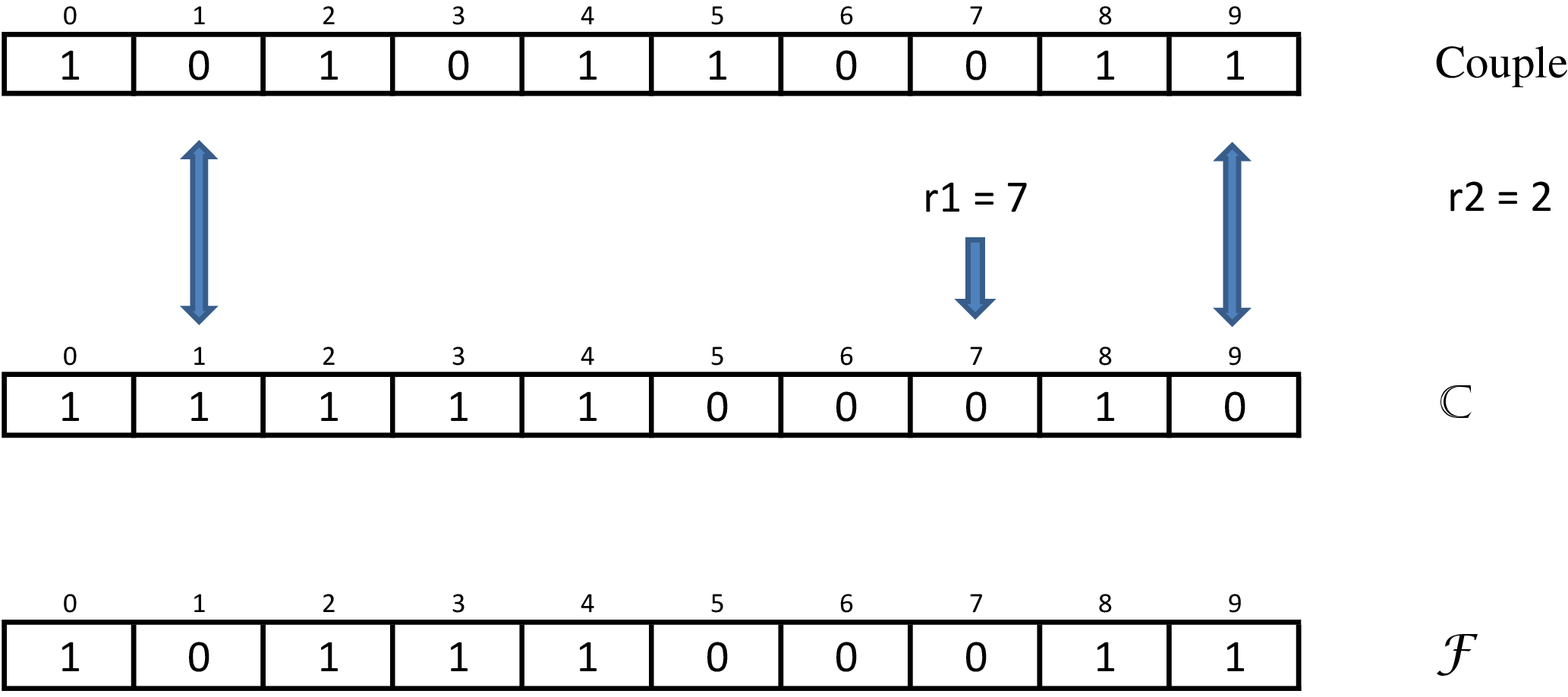}
\caption{A crossover operation between $\mathbb{C}$ and Couple to
offspring $\mathcal{F}$, $m = 10$, $p = 6$, $r1 = 7$, and $r2 = 2$.}
\end{figure}

\subsection{Mutation Operation}

\label{Mutation}

The mutation operation is based on gene/bit shifting. We use two types
of shifts: \emph{circular shift} and \emph{block shift}. In circular
shift the number of genes to be shifted and the shift direction are
randomly decided. Then, the genes are rotated accordingly. Figure~2
illustrates a three positions right circular shift mutation of chromosome $\mathbb{C}$
to offspring $\mathcal{F}$.

\begin{figure}
\label{Fig2}
\centering
\includegraphics[width=0.75\textwidth]{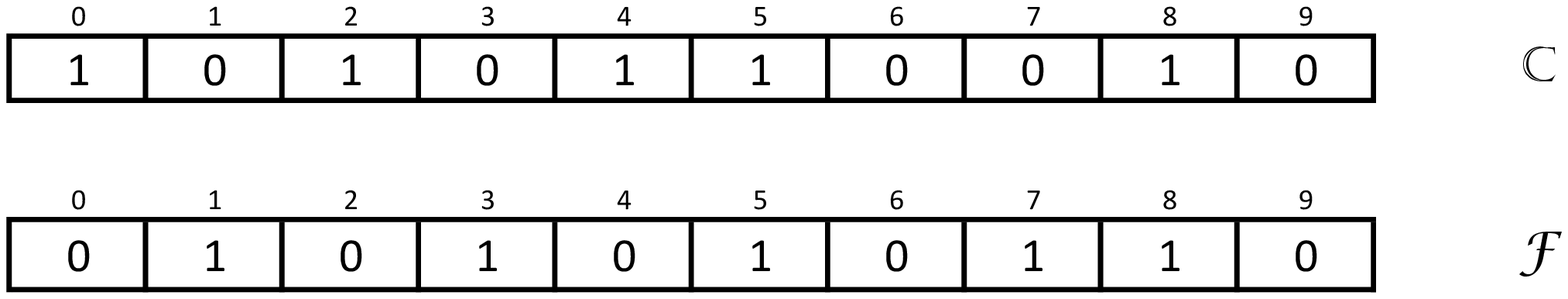}
\caption{Three positions right circular shift mutation of $\mathbb{C}$ to offspring $\mathcal{F}$.}
\end{figure}

A block shift is a circular shift on a randomly selected subsequence
of the chromosome to be mutated. The number of positions to be shifted,
the shift direction, and the subsequence indexes are randomly
decided. Then, the subsequence genes are rotated accordingly. Figure~3
shows one position left block shift on subsequence 3 to 6 of chromosome
$\mathbb{C}$ to offspring $\mathcal{F}$.
\begin{figure}
\label{Fig3}
\centering
\includegraphics[width=0.75\textwidth]{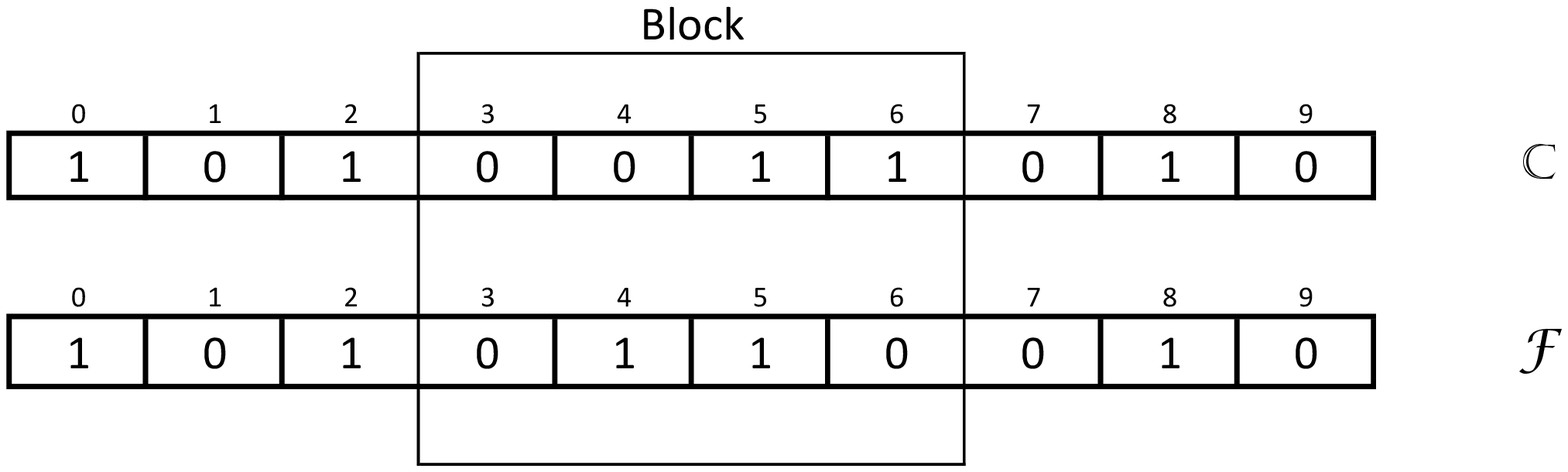}
\caption{One position left block shift mutation of $\mathbb{C}$ to offspring $\mathcal{F}$.}
\end{figure}

\subsection{Migration Operation}

Cuda does not support thread synchronizations across different blocks.
It only supports synchronizations of threads within the same block.
Therefore, we had to implement the migration operation in the Host
code. The migration takes place by selecting the best fit chromosome
computed by each block in the last Evolve kernel and adding
it to the newly generated population for the next kernel launch. Different
variations of the migration operation can be applied. For example,
the best of each block can be migrated to the same block in the next
generation, or a team of all the bests can be migrated to a single
block.

\section{Time Complexity}
\label{Complexity}

The Fitness function time complexity as indicated in
Subsection~\ref{fitness} is $O(n(m-p)) = O(nm)$.
Hence, each of the crossover and mutation cycles in the Evolve
kernel is $O(nm\ lg(NT))$. Therefore, the Evolve kernel time
complexity is
\begin{center}
$T_{E}= O(nm\ lg(NT))$.
\end{center}

The Init kernel is $O(1)$ since each thread would execute
a constant number of operations.

The complexity of generating one chromosome~(candidate solution) in the host is
$O(mp)$. Consequently, the time complexity of the first eight steps in the Host Code is 
\begin{center}
$T_S\ =\ O(nm)+ O(\mathit{NB}\ NT\ mp)$.
\par\end{center}

The remaining steps are iterative and the time complexity of a single
iteration of these steps is 
\begin{center}
$T_{H} =O(\mathit{NB}\ NT\ mp)\ +\ W$, 
\par\end{center}
where $W$~is the waiting time for the Evolve kernel to finish (Step
12).
Since $T_H$ and $T_E$ run in parallel, the algorithm total complexity is 

\begin{center}
$T_S \ + O(EvolveLimit\ \times\ max(T_H,\ T_E))$.
\par\end{center}

For maximum utilization of both Host and GPU device, $W$ must be $0$ and
$T_{H}\ =\ T_{E}$. This can be achieved by proper selection of NB and
NT within the GPU device limits. 
Further synchronization between $T_H$ and $T_E$
could be achieved by increasing the crossover and/or mutation iterations
to a number decided by an input parameter.

\section{Experimentation Results}
\label{Experimentation}

The objective of the experimentation was to test the effectiveness
of our algorithm rather than to optimize the implementation to the best
possible performance.
We have tested the algorithm on all the benchmark instances 
we had access to.
In total, we have tested 290 diversified instances
collected as follows:
40 instances from the OR-Library~\cite{Beasley},
40 instances of the so-called \emph{{}complex instances} introduced in Table 2.6
of~\cite{BorisGoldengorin2013},
and 210 instances from the discrete location problems benchmark
library~\cite{RUS-BM}.
All our experimentation were executed on
a Tesla K40 (2880 Cuda cores) hosted by HP Z820 workstation equipped
with: $2\times$ Intel Xeon processors $12$ cores each, $16$~GB
RAM, and $2\times512$~GB solid state drives. The specifications
of these equipment could be referred to in~\cite{HPZ820}\cite{Teslak40}.

The algorithm succeeded in obtaining
optimal solutions for all the 290 instances except two, namely
OR-Library pmed30 and pmed40 where a better than 99.9\% approximation
was obtained for each. The obtained results are listed in Tables~1
to~9. 
By examining these results, we can draw the following
notes and observations: 
\begin{enumerate}
\item
Our algorithm succeeded in obtaining optimal solutions for all what
so called \textquotedbl{}complex instances\textquotedbl{} as shown
in Table~9. Goldengorin et. al. introduced these forty
instances where optimal solutions for thirty of which could not be
obtained by linear programming using Elloumi formulation or pseudo
boolean formulation and data reductions~\cite{BorisGoldengorin2013}. 
\item
Our algorithm critically relies on randomization in initializing potential
solutions and enhancing them. Thus, there is no guarantee to obtain
the same results in each run of the algorithm on a given benchmark
instance. Except for OR-Library pmed30 and pmed40, nevertheless, our
implementation has shown excellent consistency in obtaining optimal
solutions over multiple runs on each tested benchmark instance, but
possibly with different timings and/or kernel counters.
The measurements listed in Tables~1
to~9 are the medians obtained from different
runs.
Furthermore, these measurements are for the kernels in which the
optimal solutions were achieved rather than for the kernels at which the
program terminated with the exceptions of pmed30 and pmed40 as
no optimal solutions were achieved.
\item
The chromosomes generation method as explained in
Subsection~\ref{generation} tremendously enhances the
candidate solutions' qualities. Moreover, the independence of
the random functions used in the Host Code and in the Evolve kernel
contributes to this enhancement as it dedicates the host random function
for generating candidate solutions. 
\item
As indicated in Subsection~\ref{fitness}, the Fitness function is
a performance bottleneck. In average, there is
one open facility in any~$\frac{m-p-1}{p} \approx \frac{m}{p}$ locations
assuming the open facilities are normally distributed.
We could have sped up this function execution by
using~$n$ threads each of which accumulates the increments of one client
($n$~threads scenario)
rather than using a single thread to accumulate the increments of
all clients (single thread scenario).
In the $n$~threads scenario, however, the Fitness function execution
time is determined by the last finishing thread,($t_l$),
whose execution time in average will be worse than $\frac{m}{p}$
and could be~$\Omega(m)$.
Each of the other~$n-1$ threads will be idle from its finishing time
till the finishing time of~$t_l$. This would
result in underutilized device cores,
and would hinder the performance scalability as~$n$ and~$m$ increase.
The single thread scenario requires more time to evaluate
the fitness of a single chromosome, but with no thread idle time.
In this scenario, the Fitness function average time is~$O(\frac{nm}{p})$.
As~$p$ scales to~$\theta(m)$, the average time could drop to~$O(n)$.
This explains the total time drop when scaling~$p$ and fixing~$n$~and~$m$
in our experimentation of the Pmed and the Complex benchmarks,
refer to Tables~1 and~9.
Evaluating the fitness of~$n$ chromosomes in the
$n$~threads scenario requires~$n^2$ threads
and~$O(m)$ average time. The same requires~$n$ threads
and~$O(n)$ average time with proper scaling of~$p$.
As the number of threads exceeds the number of available cores,
thread queuing overhead and waiting times will accumulate.
Obviously, the $n$~threads scenario requires more
threads as~$n$ scales. Thus, it is more vulnerable
to these overheads and waiting times. 
\item
The time needed to generate a population is proportional to its
size $=\ \mathit{NB} \times NT$.
We noticed that increasing the population size improves the chances of
obtaining an optimal solution, refer to Table~6 as an example.
However, determining the population
size has to be within the GPU device hardware limitations:
number of cores, threads queue depth, memory transfers, memory access
conflicts, ... etc.
\item
Increasing~$NT$ enhances the solutions' qualities as it increases
the Crossover and Mutation iterations. This could result in obtaining an
optimal solution in less number of Evolve kernel calls,
but with more time per kernel. For example, compare Tables~2 and~3.
\item
We have experimented the Crossover and Mutation impacts independently from $NT$.
In these experimentation, we determined  the number of Crossover and
Mutations iterations by an input parameter. We found that increasing the number of
iterations lead to optimal solutions in less number of kernels, but with
higher average kernel time. Tables~4 and~7 show the related results.

\item
The Migration operation impact starts from the
second Evolve kernel and onward. As per our algorithm design, the
number of chromosomes to be migrated to a next population is proportional to~$\mathit{NB}$.
Our experimentation indicated that
the number of chromosomes to be migrated from each block and their distribution
over the next kernel blocks influence the solution quality obtained by that kernel.
We experienced these impacts while testing the Pmed, Chess Board, and Large
Duality Gap-C benchmarks as they required more kernel calls than
the other benchmarks, refer to Tables~1,~3, and~8.
\item
The experimentation results are consistent with the time complexity analysis
in Section~\ref{Complexity} except for the results
shown in Tables~4 and~7 as explained above.
The Evolve kernel average time is proportional
to~$n$, $m$, and $TN$. 
Furthermore, the experimentation indicated
that this average time is also proportional to $\frac{TB \times TN}{Number\ of\ Cores}$.
This is valid because threads will be queued as
the number of threads exceeds the number of available cores.
Tables~2 and~5 show the impact of increasing~$n$ and~$m$ on the kernel time when fixing
$NT$ and $\mathit{NB}$, while Tables~2 and~6 point out the impact of increasing
$\frac{TB \times TN}{Number\ of\ Cores}$.
\end{enumerate}
\begin{table}
\label{pmed} 

\centering{}%
\resizebox{\columnwidth}{!}{
\begin{tabular}{|c|c|c|c|c|c|c|}
\hline 
Instance  & n = m  & p  & Number of  & Obtained Solution  & Number of  & Time \tabularnewline
Code  &  &  & Potential Solutions  & Approximation Ratio  & Kernel Calls  & (Sec.)\tabularnewline
\hline 
Pmed 1  & 100  & 5  & 7.53E+07  & Optimal  & 1  & 2 \tabularnewline
Pmed 2  & 100  & 10  & 1.73E+13  & Optimal  & 1  & 2 \tabularnewline
Pmed 3  & 100  & 10  & 1.73E+13  & Optimal  & 1  & 2 \tabularnewline
Pmed 4  & 100  & 20  & 5.36E+20  & Optimal  & 1  & 2 \tabularnewline
Pmed 5  & 100  & 33  & 2.95E+26  & Optimal  & 2  & 4 \tabularnewline
Pmed 6  & 200  & 5  & 2.54E+09  & Optimal  & 1  & 8 \tabularnewline
Pmed 7  & 200  & 10  & 2.25E+16  & Optimal  & 1  & 6 \tabularnewline
Pmed 8  & 200  & 20  & 1.61E+27  & Optimal  & 4  & 15 \tabularnewline
Pmed 9  & 200  & 40  & 2.05E+42  & Optimal  & 7  & 23 \tabularnewline
Pmed 10  & 200  & 67  & 1.45E+54  & Optimal  & 13  & 49 \tabularnewline
Pmed 11  & 300  & 5  & 1.96E+10  & Optimal  & 1  & 13 \tabularnewline
Pmed 12  & 300  & 10  & 1.40E+18  & Optimal  & 2  & 19 \tabularnewline
Pmed 13  & 300  & 30  & 1.73E+41  & Optimal  & 10  & 61 \tabularnewline
Pmed 14  & 300  & 60  & 9.04E+63  & Optimal  & 14  & 137 \tabularnewline
Pmed 15  & 300  & 100  & 4.16E+81  & Optimal  & 16  & 3744 \tabularnewline
Pmed 16  & 400  & 5  & 8.32E+10  & Optimal  & 1  & 23 \tabularnewline
Pmed 17  & 400  & 10  & 2.58E+19  & Optimal  & 4  & 61 \tabularnewline
Pmed 18  & 400  & 40  & 1.97E+55  & Optimal  & 15  & 158 \tabularnewline
Pmed 19  & 400  & 80  & 4.23E+85  & Optimal  & 15  & 2608 \tabularnewline
Pmed 20  & 400  & 133  & 1.26E+109  & Optimal  & 17  & 462 \tabularnewline
Pmed 21  & 500  & 5  & 2.55E+11  & Optimal  & 1  & 34 \tabularnewline
Pmed 22  & 500  & 10  & 2.46E+20  & Optimal  & 6  & 150 \tabularnewline
Pmed 23  & 500  & 50  & 2.31E+69  & Optimal  & 27  & 495 \tabularnewline
Pmed 24  & 500  & 100  & 2.04E+107  & Optimal  & 16  & 4104 \tabularnewline
Pmed 25  & 500  & 167  & 7.85E+136  & Optimal  & 14  & 2201 \tabularnewline
Pmed 26  & 600  & 5  & 6.37E+11  & Optimal  & 1  & 50 \tabularnewline
Pmed 27  & 600  & 10  & 1.55E+21  & Optimal  & 9  & 280 \tabularnewline
Pmed 28  & 600  & 60  & 2.77E+83  & Optimal  & 13  & 2918 \tabularnewline
Pmed 29  & 600  & 120  & 1.01E+129  & Optimal  & 50  & 8856 \tabularnewline
Pmed 30  & 600  & 200  & 2.51E+164  & 0.999497487  & 100  & 41687 \tabularnewline
Pmed 31  & 700  & 5  & 1.38E+12  & Optimal  & 1  & 98 \tabularnewline
Pmed 32  & 700  & 10  & 7.30E+21  & Optimal  & 3  & 186 \tabularnewline
Pmed 33  & 700  & 70  & 3.37E+97  & Optimal  & 28  & 5385 \tabularnewline
Pmed 34  & 700  & 140  & 5.03E+150  & Optimal  & 33  & 8143 \tabularnewline
Pmed 35  & 800  & 5  & 2.70E+12  & Optimal  & 2  & 249 \tabularnewline
Pmed 36  & 800  & 10  & 2.80E+22  & Optimal  & 2  & 163 \tabularnewline
Pmed 37  & 800  & 80  & 4.14E+111  & Optimal  & 15  & 8326 \tabularnewline
Pmed 38  & 900  & 5  & 4.87E+12  & Optimal  & 5  & 763 \tabularnewline
Pmed 39  & 900  & 10  & 9.14E+22  & Optimal  & 7  & 643 \tabularnewline
Pmed 40  & 900  & 90  & 5.13E+125  & 0.99980503  & 100  & 63088 \tabularnewline
\hline 
\end{tabular}
}
\caption{Results for P median benchmark instances obtained from OR-Library
with NB = 60 and NT = 256.}
\end{table}

\begin{table}
\label{pc} 

\centering{}%
\resizebox{\columnwidth}{!}{
\begin{tabular}{|c|c|c|c|c|c|c|}
\hline 
Instance  & n = m  & p  & Number of  & Obtained Solution  & Number of  & Time \tabularnewline
Code  &  &  & Potential Solutions  & Approximation Ratio  & Kernel Calls  & (Sec.)\tabularnewline
\hline 
313  & 128  & 16  & 9.3343E+19  & Optimal  & 1  & 7 \tabularnewline
323  & 128  & 16  & 9.3343E+19  & Optimal  & 1  & 7 \tabularnewline
334  & 128  & 16  & 9.3343E+19  & Optimal  & 1  & 7 \tabularnewline
434  & 128  & 16  & 9.3343E+19  & Optimal  & 1  & 7 \tabularnewline
534  & 128  & 16  & 9.3343E+19  & Optimal  & 1  & 7 \tabularnewline
634  & 128  & 16  & 9.3343E+19  & Optimal  & 2  & 12 \tabularnewline
734  & 128  & 16  & 9.3343E+19  & Optimal  & 1  & 7 \tabularnewline
834  & 128  & 16  & 9.3343E+19  & Optimal  & 1  & 7 \tabularnewline
934  & 128  & 16  & 9.3343E+19  & Optimal  & 1  & 7 \tabularnewline
1034  & 128  & 16  & 9.3343E+19  & Optimal  & 1  & 7 \tabularnewline
1134  & 128  & 16  & 9.3343E+19  & Optimal  & 1  & 7 \tabularnewline
1234  & 128  & 16  & 9.3343E+19  & Optimal  & 1  & 6 \tabularnewline
1334  & 128  & 16  & 9.3343E+19  & Optimal  & 1  & 7 \tabularnewline
1434  & 128  & 16  & 9.3343E+19  & Optimal  & 1  & 7 \tabularnewline
1534  & 128  & 16  & 9.3343E+19  & Optimal  & 1  & 7 \tabularnewline
1634  & 128  & 16  & 9.3343E+19  & Optimal  & 1  & 6 \tabularnewline
1734  & 128  & 16  & 9.3343E+19  & Optimal  & 2  & 13 \tabularnewline
1834  & 128  & 16  & 9.3343E+19  & Optimal  & 1  & 7 \tabularnewline
1934  & 128  & 16  & 9.3343E+19  & Optimal  & 1  & 7 \tabularnewline
2034  & 128  & 16  & 9.3343E+19  & Optimal  & 1  & 7 \tabularnewline
2134  & 128  & 16  & 9.3343E+19  & Optimal  & 1  & 7 \tabularnewline
2234  & 128  & 16  & 9.3343E+19  & Optimal  & 1  & 7 \tabularnewline
2334  & 128  & 16  & 9.3343E+19  & Optimal  & 1  & 7 \tabularnewline
2434  & 128  & 16  & 9.3343E+19  & Optimal  & 1  & 7 \tabularnewline
2534  & 128  & 16  & 9.3343E+19  & Optimal  & 1  & 7 \tabularnewline
2634  & 128  & 16  & 9.3343E+19  & Optimal  & 1  & 7 \tabularnewline
2734  & 128  & 16  & 9.3343E+19  & Optimal  & 1  & 7 \tabularnewline
2834  & 128  & 16  & 9.3343E+19  & Optimal  & 1  & 7 \tabularnewline
2934  & 128  & 16  & 9.3343E+19  & Optimal  & 1  & 7 \tabularnewline
3034  & 128  & 16  & 9.3343E+19  & Optimal  & 1  & 7 \tabularnewline
\hline 
\end{tabular}
}
\caption{Results for Perfect Codes Instances obtained from Discrete Location
Problems Benchmark Library with NB = 120 and NT = 256.}
\end{table}

\begin{table}
\label{CB} 
\centering{}%
\resizebox{\columnwidth}{!}{
\begin{tabular}{|c|c|c|c|c|c|c|}
\hline 
Instance  & n = m  & p  & Number of  & Obtained Solution  & Number of  & Time \tabularnewline
Code  &  &  & Potential Solutions  & Approximation Ratio  & Kernel Calls  & (Sec.)\tabularnewline
\hline 
334  & 144  & 16  & 6.88E+20  & Optimal  & 11  & 36 \tabularnewline
434  & 144  & 16  & 6.88E+20  & Optimal  & 12  & 39 \tabularnewline
534  & 144  & 16  & 6.88E+20  & Optimal  & 11  & 59 \tabularnewline
634  & 144  & 16  & 6.88E+20  & Optimal  & 14  & 45 \tabularnewline
734  & 144  & 16  & 6.88E+20  & Optimal  & 7  & 25 \tabularnewline
834  & 144  & 16  & 6.88E+20  & Optimal  & 9  & 49 \tabularnewline
934  & 144  & 16  & 6.88E+20  & Optimal  & 13  & 42 \tabularnewline
1034  & 144  & 16  & 6.88E+20  & Optimal  & 29  & 100 \tabularnewline
1134  & 144  & 16  & 6.88E+20  & Optimal  & 7  & 24 \tabularnewline
1234  & 144  & 16  & 6.88E+20  & Optimal  & 21  & 67 \tabularnewline
1334  & 144  & 16  & 6.88E+20  & Optimal  & 8  & 29 \tabularnewline
1434  & 144  & 16  & 6.88E+20  & Optimal  & 10  & 34 \tabularnewline
1534  & 144  & 16  & 6.88E+20  & Optimal  & 19  & 5 \tabularnewline
1634  & 144  & 16  & 6.88E+20  & Optimal  & 18  & 5 \tabularnewline
1734  & 144  & 16  & 6.88E+20  & Optimal  & 18  & 56 \tabularnewline
1834  & 144  & 16  & 6.88E+20  & Optimal  & 46  & 142 \tabularnewline
1934  & 144  & 16  & 6.88E+20  & Optimal  & 8  & 44 \tabularnewline
2034  & 144  & 16  & 6.88E+20  & Optimal  & 11  & 37 \tabularnewline
2134  & 144  & 16  & 6.88E+20  & Optimal  & 14  & 44 \tabularnewline
2234  & 144  & 16  & 6.88E+20  & Optimal  & 22  & 75 \tabularnewline
2334  & 144  & 16  & 6.88E+20  & Optimal  & 13  & 69 \tabularnewline
2434  & 144  & 16  & 6.88E+20  & Optimal  & 7  & 23 \tabularnewline
2534  & 144  & 16  & 6.88E+20  & Optimal  & 11  & 35 \tabularnewline
2634  & 144  & 16  & 6.88E+20  & Optimal  & 2  & 9 \tabularnewline
2734  & 144  & 16  & 6.88E+20  & Optimal  & 6  & 23 \tabularnewline
2834  & 144  & 16  & 6.88E+20  & Optimal  & 6  & 20 \tabularnewline
2934  & 144  & 16  & 6.88E+20  & Optimal  & 1  & 6 \tabularnewline
3034  & 144  & 16  & 6.88E+20  & Optimal  & 17  & 54 \tabularnewline
3134  & 144  & 16  & 6.88E+20  & Optimal  & 11  & 37 \tabularnewline
3234  & 144  & 16  & 6.88E+20  & Optimal  & 10  & 33 \tabularnewline
\hline 
\end{tabular}
}
\caption{Results for Chess Board Instances obtained from Discrete Location
Problems Benchmark Library with NB = 480 and NT = 96.}
\end{table}

\begin{table}
\label{FPPK11} 

\centering{}%
\resizebox{\columnwidth}{!}{
\begin{tabular}{|c|c|c|c|c|c|c|}
\hline 
Instance  & n = m  & p  & Number of  & Obtained Solution  & Number of  & Time \tabularnewline
Code  &  &  & Potential Solutions  & Approximation Ratio  & Kernel Calls  & (Sec.)\tabularnewline
\hline 
1  & 133  & 12  & 3.84E+16  & Optimal  & 1  & 78 \tabularnewline
2  & 133  & 12  & 3.84E+16  & Optimal  & 1  & 78 \tabularnewline
3  & 133  & 12  & 3.84E+16  & Optimal  & 1  & 81 \tabularnewline
4  & 133  & 12  & 3.84E+16  & Optimal  & 1  & 78 \tabularnewline
5  & 133  & 12  & 3.84E+16  & Optimal  & 1  & 76 \tabularnewline
6  & 133  & 12  & 3.84E+16  & Optimal  & 1  & 79 \tabularnewline
7  & 133  & 12  & 3.84E+16  & Optimal  & 1  & 75 \tabularnewline
8  & 133  & 12  & 3.84E+16  & Optimal  & 1  & 76 \tabularnewline
9  & 133  & 12  & 3.84E+16  & Optimal  & 1  & 76 \tabularnewline
10  & 133  & 12  & 3.84E+16  & Optimal  & 1  & 78 \tabularnewline
11  & 133  & 12  & 3.84E+16  & Optimal  & 1  & 75 \tabularnewline
12  & 133  & 12  & 3.84E+16  & Optimal  & 1  & 76 \tabularnewline
13  & 133  & 12  & 3.84E+16  & Optimal  & 1  & 76 \tabularnewline
14  & 133  & 12  & 3.84E+16  & Optimal  & 1  & 81 \tabularnewline
15  & 133  & 12  & 3.84E+16  & Optimal  & 1  & 76 \tabularnewline
16  & 133  & 12  & 3.84E+16  & Optimal  & 1  & 77 \tabularnewline
17  & 133  & 12  & 3.84E+16  & Optimal  & 1  & 75 \tabularnewline
18  & 133  & 12  & 3.84E+16  & Optimal  & 1  & 80 \tabularnewline
19  & 133  & 12  & 3.84E+16  & Optimal  & 1  & 75 \tabularnewline
20  & 133  & 12  & 3.84E+16  & Optimal  & 1  & 77 \tabularnewline
21  & 133  & 12  & 3.84E+16  & Optimal  & 1  & 77 \tabularnewline
22  & 133  & 12  & 3.84E+16  & Optimal  & 1  & 75 \tabularnewline
23  & 133  & 12  & 3.84E+16  & Optimal  & 1  & 77 \tabularnewline
24  & 133  & 12  & 3.84E+16  & Optimal  & 1  & 74 \tabularnewline
25  & 133  & 12  & 3.84E+16  & Optimal  & 1  & 78 \tabularnewline
26  & 133  & 12  & 3.84E+16  & Optimal  & 1  & 74 \tabularnewline
27  & 133  & 12  & 3.84E+16  & Optimal  & 1  & 78 \tabularnewline
28  & 133  & 12  & 3.84E+16  & Optimal  & 1  & 79 \tabularnewline
29  & 133  & 12  & 3.84E+16  & Optimal  & 1  & 73 \tabularnewline
30  & 133  & 12  & 3.84E+16  & Optimal  & 1  & 76 \tabularnewline
\hline 
\end{tabular}
}
\caption{Results for Finite Projective Planes Instances, K = 11 obtained from
Discrete Location Problems Benchmark Library with NB = 120 and NT
= 256. In this experiment, the number of crossover and mutation iterations were preset
to~60.}
\end{table}

\begin{table}
\label{FPPK17} 

\centering{}%
\resizebox{\columnwidth}{!}{
\begin{tabular}{|c|c|c|c|c|c|c|}
\hline 
Instance  & n = m  & p  & Number of  & Obtained Solution  & Number of  & Time \tabularnewline
Code  &  &  & Potential Solutions  & Approximation Ratio  & Kernel Calls  & (Sec.)\tabularnewline
\hline 
1  & 307  & 18  & 5.51E+28  & Optimal  & 1  & 39 \tabularnewline
2  & 307  & 18  & 5.51E+28  & Optimal  & 1  & 40 \tabularnewline
3  & 307  & 18  & 5.51E+28  & Optimal  & 1  & 40 \tabularnewline
4  & 307  & 18  & 5.51E+28  & Optimal  & 1  & 40 \tabularnewline
5  & 307  & 18  & 5.51E+28  & Optimal  & 1  & 40 \tabularnewline
6  & 307  & 18  & 5.51E+28  & Optimal  & 6  & 212 \tabularnewline
7  & 307  & 18  & 5.51E+28  & Optimal  & 1  & 39 \tabularnewline
8  & 307  & 18  & 5.51E+28  & Optimal  & 1  & 39 \tabularnewline
9  & 307  & 18  & 5.51E+28  & Optimal  & 1  & 39 \tabularnewline
10  & 307  & 18  & 5.51E+28  & Optimal  & 1  & 39 \tabularnewline
11  & 307  & 18  & 5.51E+28  & Optimal  & 1  & 38 \tabularnewline
12  & 307  & 18  & 5.51E+28  & Optimal  & 2  & 73 \tabularnewline
13  & 307  & 18  & 5.51E+28  & Optimal  & 2  & 73 \tabularnewline
14  & 307  & 18  & 5.51E+28  & Optimal  & 1  & 39 \tabularnewline
15  & 307  & 18  & 5.51E+28  & Optimal  & 2  & 74 \tabularnewline
16  & 307  & 18  & 5.51E+28  & Optimal  & 2  & 71 \tabularnewline
17  & 307  & 18  & 5.51E+28  & Optimal  & 1  & 39 \tabularnewline
18  & 307  & 18  & 5.51E+28  & Optimal  & 1  & 39 \tabularnewline
19  & 307  & 18  & 5.51E+28  & Optimal  & 1  & 39 \tabularnewline
20  & 307  & 18  & 5.51E+28  & Optimal  & 1  & 40 \tabularnewline
21  & 307  & 18  & 5.51E+28  & Optimal  & 1  & 39 \tabularnewline
22  & 307  & 18  & 5.51E+28  & Optimal  & 5  & 178 \tabularnewline
23  & 307  & 18  & 5.51E+28  & Optimal  & 4  & 140 \tabularnewline
24  & 307  & 18  & 5.51E+28  & Optimal  & 1  & 40 \tabularnewline
25  & 307  & 18  & 5.51E+28  & Optimal  & 1  & 39 \tabularnewline
26  & 307  & 18  & 5.51E+28  & Optimal  & 1  & 38 \tabularnewline
27  & 307  & 18  & 5.51E+28  & Optimal  & 2  & 75 \tabularnewline
28  & 307  & 18  & 5.51E+28  & Optimal  & 1  & 40 \tabularnewline
29  & 307  & 18  & 5.51E+28  & Optimal  & 1  & 41 \tabularnewline
30  & 307  & 18  & 5.51E+28  & Optimal  & 1  & 39 \tabularnewline
\hline 
\end{tabular}
}
\caption{Results for Finite Projective Planes Instances, K = 17 obtained from
Discrete Location Problems Benchmark Library with NB = 120 and NT
= 256.}
\end{table}

\begin{table}
\label{GAPA} 
\centering{}%
\resizebox{\columnwidth}{!}{
\begin{tabular}{|c|c|c|c|c|c|c|}
\hline 
Instance  & n = m  & p  & Number of  & Obtained Solution  & Number of  & Time \tabularnewline
Code  &  &  & Potential Solutions  & Approximation Ratio  & Kernel Calls  & (Sec.)\tabularnewline
\hline 
332  & 100  & 12  & 1.05E+15  & Optimal  & 1  & 46 \tabularnewline
432  & 100  & 12  & 1.05E+15  & Optimal  & 2  & 85 \tabularnewline
532  & 100  & 12  & 1.05E+15  & Optimal  & 2  & 86 \tabularnewline
632  & 100  & 12  & 1.05E+15  & Optimal  & 1  & 46 \tabularnewline
732  & 100  & 12  & 1.05E+15  & Optimal  & 1  & 46 \tabularnewline
832  & 100  & 12  & 1.05E+15  & Optimal  & 1  & 45 \tabularnewline
932  & 100  & 12  & 1.05E+15  & Optimal  & 1  & 45 \tabularnewline
1032  & 100  & 12  & 1.05E+15  & Optimal  & 1  & 60 \tabularnewline
1132  & 100  & 12  & 1.05E+15  & Optimal  & 1  & 60 \tabularnewline
1232  & 100  & 12  & 1.05E+15  & Optimal  & 1  & 52 \tabularnewline
1332  & 100  & 12  & 1.05E+15  & Optimal  & 1  & 46 \tabularnewline
1432  & 100  & 12  & 1.05E+15  & Optimal  & 1  & 47 \tabularnewline
1532  & 100  & 12  & 1.05E+15  & Optimal  & 2  & 91 \tabularnewline
1632  & 100  & 12  & 1.05E+15  & Optimal  & 1  & 48 \tabularnewline
1732  & 100  & 12  & 1.05E+15  & Optimal  & 1  & 49 \tabularnewline
1832  & 100  & 12  & 1.05E+15  & Optimal  & 7  & 304 \tabularnewline
1932  & 100  & 12  & 1.05E+15  & Optimal  & 1  & 48 \tabularnewline
2032  & 100  & 12  & 1.05E+15  & Optimal  & 1  & 48 \tabularnewline
2132  & 100  & 12  & 1.05E+15  & Optimal  & 1  & 49 \tabularnewline
2232  & 100  & 12  & 1.05E+15  & Optimal  & 1  & 49 \tabularnewline
2332  & 100  & 12  & 1.05E+15  & Optimal  & 1  & 49 \tabularnewline
2432  & 100  & 12  & 1.05E+15  & Optimal  & 1  & 49 \tabularnewline
2532  & 100  & 12  & 1.05E+15  & Optimal  & 1  & 49 \tabularnewline
2632  & 100  & 12  & 1.05E+15  & Optimal  & 2  & 91 \tabularnewline
2732  & 100  & 12  & 1.05E+15  & Optimal  & 25  & 1027 \tabularnewline
2832  & 100  & 12  & 1.05E+15  & Optimal  & 1  & 49 \tabularnewline
2932  & 100  & 12  & 1.05E+15  & Optimal  & 1  & 45 \tabularnewline
3032  & 100  & 12  & 1.05E+15  & Optimal  & 1  & 45 \tabularnewline
3132  & 100  & 12  & 1.05E+15  & Optimal  & 1  & 43 \tabularnewline
3232  & 100  & 12  & 1.05E+15  & Optimal  & 2  & 88 \tabularnewline
\hline 
\end{tabular}
}
\caption{Results for Large Duality Gap-A Instances obtained from Discrete Location
Problems Benchmark Library with NB = 1500 and NT = 64.}
\end{table}

\begin{table}
\label{GAPB} 

\centering{}%
\resizebox{\columnwidth}{!}{
\begin{tabular}{|c|c|c|c|c|c|c|}
\hline 
Instance  & n = m  & p  & Number of  & Obtained Solution  & Number of  & Time \tabularnewline
Code  &  &  & Potential Solutions  & Approximation Ratio  & Kernel Calls  & (Sec.)\tabularnewline
\hline 
331  & 100  & 14  & 4.42E+16  & Optimal  & 2  & 54 \tabularnewline
431  & 100  & 14  & 4.42E+16  & Optimal  & 2  & 54 \tabularnewline
531  & 100  & 14  & 4.42E+16  & Optimal  & 2  & 52 \tabularnewline
631  & 100  & 14  & 4.42E+16  & Optimal  & 2  & 58 \tabularnewline
731  & 100  & 14  & 4.42E+16  & Optimal  & 5  & 130 \tabularnewline
831  & 100  & 14  & 4.42E+16  & Optimal  & 1  & 27 \tabularnewline
931  & 100  & 14  & 4.42E+16  & Optimal  & 2  & 58 \tabularnewline
1031  & 100  & 14  & 4.42E+16  & Optimal  & 1  & 30 \tabularnewline
1131  & 100  & 14  & 4.42E+16  & Optimal  & 1  & 30 \tabularnewline
1231  & 100  & 14  & 4.42E+16  & Optimal  & 6  & 156 \tabularnewline
1331  & 100  & 14  & 4.42E+16  & Optimal  & 1  & 29 \tabularnewline
1431  & 100  & 14  & 4.42E+16  & Optimal  & 5  & 138 \tabularnewline
1531  & 100  & 14  & 4.42E+16  & Optimal  & 6  & 171 \tabularnewline
1631  & 100  & 14  & 4.42E+16  & Optimal  & 2  & 52 \tabularnewline
1731  & 100  & 14  & 4.42E+16  & Optimal  & 1  & 30 \tabularnewline
1831  & 100  & 14  & 4.42E+16  & Optimal  & 2  & 52 \tabularnewline
1931  & 100  & 14  & 4.42E+16  & Optimal  & 1  & 26 \tabularnewline
2031  & 100  & 14  & 4.42E+16  & Optimal  & 1  & 27 \tabularnewline
2131  & 100  & 14  & 4.42E+16  & Optimal  & 1  & 26 \tabularnewline
2231  & 100  & 14  & 4.42E+16  & Optimal  & 2  & 62 \tabularnewline
2331  & 100  & 14  & 4.42E+16  & Optimal  & 1  & 28 \tabularnewline
2431  & 100  & 14  & 4.42E+16  & Optimal  & 5  & 132 \tabularnewline
2531  & 100  & 14  & 4.42E+16  & Optimal  & 1  & 27 \tabularnewline
2631  & 100  & 14  & 4.42E+16  & Optimal  & 10  & 252 \tabularnewline
2731  & 100  & 14  & 4.42E+16  & Optimal  & 2  & 61 \tabularnewline
2831  & 100  & 14  & 4.42E+16  & Optimal  & 1  & 32 \tabularnewline
2931  & 100  & 14  & 4.42E+16  & Optimal  & 15  & 387 \tabularnewline
3031  & 100  & 14  & 4.42E+16  & Optimal  & 1  & 26 \tabularnewline
3131  & 100  & 14  & 4.42E+16  & Optimal  & 1  & 30 \tabularnewline
3231  & 100  & 14  & 4.42E+16  & Optimal  & 3  & 75 \tabularnewline
\hline 
\end{tabular}
}
\caption{Results for Large Duality Gap-B Instances obtained from Discrete Location
Problems Benchmark Library with NB = 120 and NT = 256. In this experiment,
the number of crossover and mutation iterations was preset to~20.}
\end{table}

\begin{table}
\label{GAPC} 

\centering{}%
\resizebox{\columnwidth}{!}{
\begin{tabular}{|c|c|c|c|c|c|c|}
\hline 
Instance  & n = m  & p  & Number of  & Obtained Solution  & Number of  & Time \tabularnewline
Code  &  &  & Potential Solutions  & Approximation Ratio  & Kernel Calls  & (Sec.)\tabularnewline
\hline 
333  & 100  & 14  & 4.42E+16  & Optimal  & 3  & 37 \tabularnewline
433  & 100  & 14  & 4.42E+16  & Optimal  & 2  & 26 \tabularnewline
533  & 100  & 14  & 4.42E+16  & Optimal  & 2  & 26 \tabularnewline
633  & 100  & 14  & 4.42E+16  & Optimal  & 4  & 47 \tabularnewline
733  & 100  & 14  & 4.42E+16  & Optimal  & 3  & 38 \tabularnewline
833  & 100  & 14  & 4.42E+16  & Optimal  & 3  & 37 \tabularnewline
933  & 100  & 14  & 4.42E+16  & Optimal  & 2  & 26 \tabularnewline
1033  & 100  & 14  & 4.42E+16  & Optimal  & 4  & 49 \tabularnewline
1133  & 100  & 14  & 4.42E+16  & Optimal  & 7  & 80 \tabularnewline
1233  & 100  & 14  & 4.42E+16  & Optimal  & 11  & 126 \tabularnewline
1333  & 100  & 14  & 4.42E+16  & Optimal  & 20  & 218 \tabularnewline
1433  & 100  & 14  & 4.42E+16  & Optimal  & 2  & 27 \tabularnewline
1533  & 100  & 14  & 4.42E+16  & Optimal  & 32  & 363 \tabularnewline
1633  & 100  & 14  & 4.42E+16  & Optimal  & 3  & 37 \tabularnewline
1733  & 100  & 14  & 4.42E+16  & Optimal  & 2  & 26 \tabularnewline
1833  & 100  & 14  & 4.42E+16  & Optimal  & 18  & 198 \tabularnewline
1933  & 100  & 14  & 4.42E+16  & Optimal  & 3  & 36 \tabularnewline
2033  & 100  & 14  & 4.42E+16  & Optimal  & 5  & 59 \tabularnewline
2133  & 100  & 14  & 4.42E+16  & Optimal  & 7  & 81 \tabularnewline
2233  & 100  & 14  & 4.42E+16  & Optimal  & 2  & 26 \tabularnewline
2333  & 100  & 14  & 4.42E+16  & Optimal  & 4  & 47 \tabularnewline
2433  & 100  & 14  & 4.42E+16  & Optimal  & 9  & 105 \tabularnewline
2533  & 100  & 14  & 4.42E+16  & Optimal  & 22  & 245 \tabularnewline
2633  & 100  & 14  & 4.42E+16  & Optimal  & 3  & 37 \tabularnewline
2733  & 100  & 14  & 4.42E+16  & Optimal  & 78  & 873 \tabularnewline
2833  & 100  & 14  & 4.42E+16  & Optimal  & 2  & 27 \tabularnewline
2933  & 100  & 14  & 4.42E+16  & Optimal  & 8  & 92 \tabularnewline
3033  & 100  & 14  & 4.42E+16  & Optimal  & 1  & 15 \tabularnewline
3133  & 100  & 14  & 4.42E+16  & Optimal  & 2  & 26 \tabularnewline
3233  & 100  & 14  & 4.42E+16  & Optimal  & 3  & 37 \tabularnewline
\hline 
\end{tabular}
}
\caption{Results for Large Duality Gap-C Instances obtained from Discrete Location
Problems Benchmark Library with NB = 1500 and NT = 32.}
\end{table}

\begin{table}
\label{Complex} 

\centering{}%
\resizebox{\columnwidth}{!}{
\begin{tabular}{|c|c|c|c|c|c|}
\hline 
n = m  & p  & Number of  & Obtained Solution  & Number of  & Time \tabularnewline
 &  & Potential Solutions  & Approximation Ratio  & Kernel Calls  & (Sec.)\tabularnewline
\hline 
100  & 5  & 7.53E+07  & Optimal  & 1  & 10 \tabularnewline
100  & 10  & 1.73E+13  & Optimal  & 1  & 11 \tabularnewline
100  & 20  & 5.36E+20  & Optimal  & 1  & 11 \tabularnewline
100  & 33  & 2.95E+26  & Optimal  & 1  & 9 \tabularnewline
200  & 5  & 2.54E+09  & Optimal  & 1  & 41 \tabularnewline
200  & 10  & 2.25E+16  & Optimal  & 1  & 39 \tabularnewline
200  & 20  & 1.61E+27  & Optimal  & 1  & 40 \tabularnewline
200  & 40  & 2.05E+42  & Optimal  & 1  & 62 \tabularnewline
200  & 67  & 1.45E+54  & Optimal  & 1  & 50 \tabularnewline
300  & 5  & 1.96E+10  & Optimal  & 1  & 86 \tabularnewline
300  & 10  & 1.40E+18  & Optimal  & 1  & 143 \tabularnewline
300  & 30  & 1.73E+41  & Optimal  & 1  & 83 \tabularnewline
300  & 60  & 9.04E+63  & Optimal  & 1  & 95 \tabularnewline
300  & 100  & 4.16E+81  & Optimal  & 1  & 111 \tabularnewline
400  & 5  & 8.32E+10  & Optimal  & 1  & 256 \tabularnewline
400  & 10  & 2.58E+19  & Optimal  & 1  & 253 \tabularnewline
400  & 40  & 1.97E+55  & Optimal  & 1  & 147 \tabularnewline
400  & 80  & 4.23E+85  & Optimal  & 1  & 172 \tabularnewline
400  & 133  & 1.26E+109  & Optimal  & 1  & 197 \tabularnewline
500  & 5  & 2.55E+11  & Optimal  & 1  & 401 \tabularnewline
500  & 10  & 2.46E+20  & Optimal  & 1  & 389 \tabularnewline
500  & 50  & 2.31E+69  & Optimal  & 2  & 430 \tabularnewline
500  & 100  & 2.04E+107  & Optimal  & 2  & 486 \tabularnewline
500  & 167  & 7.85E+136  & Optimal  & 1  & 313 \tabularnewline
600  & 5  & 6.37E+11  & Optimal  & 1  & 579 \tabularnewline
600  & 10  & 1.55E+21  & Optimal  & 2  & 604 \tabularnewline
600  & 60  & 2.77E+83  & Optimal  & 2  & 617 \tabularnewline
600  & 120  & 1.01E+129  & Optimal  & 2  & 706 \tabularnewline
600  & 200  & 2.51E+164  & Optimal  & 1  & 453 \tabularnewline
700  & 5  & 1.38E+12  & Optimal  & 2  & 846 \tabularnewline
700  & 10  & 7.30E+21  & Optimal  & 2  & 818 \tabularnewline
700  & 70  & 3.37E+97  & Optimal  & 2  & 839 \tabularnewline
700  & 140  & 5.03E+150  & Optimal  & 2  & 960 \tabularnewline
800  & 5  & 2.70E+12  & Optimal  & 3  & 1660 \tabularnewline
800  & 10  & 2.80E+22  & Optimal  & 2  & 1507 \tabularnewline
800  & 80  & 4.14E+111  & Optimal  & 2  & 1123 \tabularnewline
900  & 5  & 4.87E+12  & Optimal  & 1  & 1291 \tabularnewline
900  & 10  & 9.14E+22  & Optimal  & 2  & 1905 \tabularnewline
900  & 90  & 5.13E+125  & Optimal  & 2  & 1407 \tabularnewline
\hline 
\end{tabular}
}
\caption{Results for the Complex Instances Introduced in~\cite{BorisGoldengorin2013},
where NB = 120 and NT = 256.}
\end{table}
\section{Conclusions}
In this paper, we present a new genetic algorithm for the PMP based on
GPU and pseudo-Boolean formulation.
The algorithm is composed of Host code and Device code.
The host randomly generates a population of chromosomes (candidate solutions) and passes them to
a device kernel for fitness evaluations and enhancements.
This basically iterates
with migrating the best fit chromosomes from the current population
to the next. The algorithm terminates at reaching an iteration limit
or over saturating the solution enhancement.
The algorithm is implemented using Cuda C version~7.5, and it was tested on
290 different benchmark instances.
It has succeeded in obtaining optimal solutions for all the 290 instances except
two for which better than 99.9\% approximations have been obtained.

There are several venues for our future work on this topic.
First, we will be working on identifying and developing
solution enhancement operations that shall improve our algorithm
performance.
Second, we shall analyze and experiment the algorithm scalability limits on
different GPGPU platforms.
Third, we will investigate applying the presented algorithm on different
variations of the facility location problems.
\bibliographystyle{plain}
\bibliography{bibfile}
\end{document}